\documentclass[11pt]{article}
\usepackage{moriond,epsfig,amsmath}

\def\Journal#1#2#3#4{{#1} {\bf #2}, #3 (#4)}

\def\NPB{{\em Nucl. Phys.} B}
\def\PLB{{\em Phys. Lett.}  B}

\def\PRD{{\em Phys. Rev.} D}

\def\be{\begin{equation}}
\def\ee{\end{equation}}
\def\bea{\begin{eqnarray}}
\def\eea{\end{eqnarray}}

\begin{document}
\vspace*{4cm}
\title{CLASSICAL RUNNING AND SYMMETRY BREAKING IN MODELS WITH TWO EXTRA DIMENSIONS}

\author{ C. PAPINEAU }

\address{Laboratoire de Physique Th\'eorique, Universit\'e Orsay-Paris Sud\\
91405 Orsay Cedex, France \\ Centre de Physique Th\'eorique, \'Ecole Polytechnique\\ 91128 Palaiseau Cedex, France}

\maketitle\abstracts{
We consider a codimension two scalar theory with brane-localised Higgs type potential. The six-dimensional field has Dirichlet boundary condition on the bounds of the transverse compact space. The regularisation of the brane singularity yields renormalisation group evolution for the localised couplings at the classical level. In particular, a tachyonic mass term grows at large distances and hits a Landau pole. We exhibit a peculiar value of the bare coupling such that the running mass parameter becomes large precisely at the compactification scale, and the effective four-dimensional zero mode is massless. Above the critical coupling, spontaneous symmetry breaking occurs and there is a very light state.}

\section{Introduction}

Dynamical generation of small mass scales via logarithmic
renormalisation group running often occurs
in theories which, in the limit of vanishing couplings, possess
massless degrees of freedom. Once small couplings are turned on in
UV, a theory may become strongly coupled in IR, with the
corresponding IR scale serving as the mass scale of the low energy
effective theory. A notable example of such a case is of course QCD.

In these proceedings, we present a class of models in which massless or very light degrees of freedom
emerge upon turning on small couplings in a theory which originally
had heavy states only. 
Unless the existence of massless states is dictated by a symmetry
(Goldstone modes), this requires some sort of IR--UV mixing. In general terms, our models have two widely separated
high energy scales, the UV cutoff scale $\Lambda$ and the intermediate
compactification scale $R^{-1}$. Once the small coupling $\mu = \mu (\Lambda)$ is turned
on at the UV cutoff scale, it experiences renormalisation group
running and increases towards low energies. Remarkably enough, the coupling becomes stronger in the infrared provided that its bare value is negative or, equivalently, the 4d localised mass is tachyonic. There exists a small critical value $\mu(\Lambda) = \mu_c$ such that the running
coupling blows up precisely at the scale $R^{-1}$, i.e.,
$\mu(R^{-1}) = \infty$. At this point a massless mode appears in the
spectrum.

For $\mu $ slightly smaller than $\mu_c$, the light mode has
positive mass squared proportional to $(\mu_c^2 - \mu^2)$, whereas
at $\mu$ slightly above $\mu_c$ the low energy theory is in a
symmetry breaking phase. Thus, $\mu(\Lambda) = \mu_c$ is the point
of the second order phase transition. Let us stress that these features occur at the level of classical field theory.

In our models as they stand, the bare couplings are free parameters, so the choice
$\mu \approx \mu_c$ (and hence the theory near the phase transition point) requires fine
tuning. It would be interesting to see if there exists a
dynamical mechanism driving the theory to the critical coupling (more generally, driving
the parameters to the point where $\mu (R^{-1}) = \infty$).

\section{The model}

We consider a scalar field in a six-dimensional flat space. There is no scalar potential in the bulk, and Higgs-type scalar potential on a brane located at the origin of the compact dimensions. The action is~\footnote{We  use the
convention $(+,-,-,-,-,-)$ for the metric.}
\begin{equation}
\begin{array}{ccc}
&& \displaystyle{S} \ = \ \displaystyle{\int d^4 x d^2 y \ \left[ {1 \over 2} (\partial_M \phi)^2 \
- \  V_\delta (\phi) \right]} \ ,  \\
&& \displaystyle{V_\delta (\phi)} \ = \ \displaystyle{\left(-\frac{\mu^2}{2}\phi^2 + \frac{\lambda}{4}
\phi^4 \right) \cdot \delta^2 (y)}  \ . 
\end{array}
\label{1}
\end{equation}

The singularity at ${ y}=0$ is resolved by introducing a disk $r \leq \epsilon$ and regularising the potential into
\begin{equation}
\begin{array}{ccc}
&& V(\phi) \ = \ \displaystyle{\frac{1}{\pi \epsilon^2}\left(-\frac{\mu^2}{2}\phi^2 +
\frac{\lambda}{4} \phi^4 \right)}  \quad {\rm for} \ 0 \leq r \leq \epsilon \ ,   \\
&& V(\phi) \ = \ 0  \quad \quad \quad \quad \quad \quad \quad \quad \quad \quad
{\rm for} \ \epsilon  \leq r \leq R \ . 
\end{array}
\label{potential}
\end{equation}

We consider a theory on a disk whose radius $R$ is very large, $R^{-1} \ll \epsilon^{-1}$. The parameter $\mu^2$ is dimensionless and may be considered a very small coupling constant i.e. $\mu^2 \ll 1$.
Goldberger and Wise~\cite{gw} argued that the coupling runs~\footnote{The
classical running was recently discussed in connection with neutrino
masses~\cite{dgv}. The phenomenon is well-known in two-dimensional
quantum mechanics~\cite{qm}.} with
the energy scale $Q$ as
\begin{equation}
\mu^2 (Q) = \frac{\mu^2}{1 + \frac{\mu^2}{2\pi}
\ln \frac{Q}{\Lambda}} \quad ,
\label{running}
\end{equation}
where $\Lambda$ is the UV cutoff and $ \mu^2 \equiv \mu^2(\Lambda)$ is the coupling constant entering the potential (\ref{potential}). This regularisation has a natural interpretation as finite transverse size $\epsilon=\Lambda^{-1}$ of the brane. Equation (\ref{running}) implies that the coupling $\mu^2(Q)$ grows in the infrared.

Let us impose the Dirichlet boundary condition
\[
   \phi(r=R) \ = \ 0 \ .
\]
If the coupling vanishes, or, more generally, if
the running coupling at the compactification scale is small,
$\mu^2(R^{-1}) \ll 1$, the theory is fully six-dimensional: there are no
zero or light modes, whereas the Kaluza-Klein states have masses
of order $R^{-1}$.

The question is what happens when $\mu^2 (R^{-1})$ hits the infrared
pole. In other words, what does this theory describe when the bare coupling $\mu^2$ is equal or
close to its critical value
\begin{equation}
    \mu_c^2 \ = \ \frac{2\pi}{\ln \frac{R}{\epsilon}} \ .
\label{muc}
\end{equation}
As follows from this expression, $\mu_c^2$ can (and actually, for perturbative treatment, has to) be small, which
requires that $\ln ( R / \epsilon )$ is large. Thus, it is legitimate to make use of the leading-log approximation and this is
what we are going to do. Incidentally, in the theory on a disk, the
formula for the running (\ref{running}) is also valid in the leading-log approximation only.

Splitting the four-dimensional usual coordinates and the transverse ones, the 6d equations of motion can be read from the regularised action (\ref{potential})
\begin{equation}
\left\{
\begin{array}{lcc}
-p^2 \phi - \triangle_2 \phi + \displaystyle{\left[-\frac{\mu^2}{\pi \epsilon^2} \phi + \frac{\lambda}{\pi \epsilon^2} \phi^3\right]} \ = \ 0 \quad  \ {\rm for} \ 0 \leq r \leq \epsilon \ , \\
-p^2 \phi - \triangle_2 \phi \ = \ 0 \quad \quad \quad \quad \quad \quad \quad \quad \quad \quad  \ {\rm for} \ \epsilon \leq r \leq R \ , \\
\end{array}
\right.
\label{eom}
\end{equation}
where $p^2 \phi$ is to be understood as the 4d momentum. These equations are indeed Schr\"odinger equations in two dimensions.

In the following section, we present some results when fine-tuning $\mu^2(\Lambda)$ around $\mu_c^2$.

\section{Six-dimensional description\label{6d}}

Let us first work above the background $\phi=0$ so that we can consider the linearised equations of motion. We decompose the scalar field into $\phi(x^{\mu},r,\theta)=e^{il\theta}\phi_l(r)\sigma_l(x^{\mu})$ with $l$ an integer. With this set, $\sigma_0(x^{\mu})$ is the usual zero mode that one would have derived from a Kaluza-Klein scheme when working with the torus. In the following, we shall only consider this mode.

Plugging the previous decomposition into the equations of motion (\ref{eom}) and asking that the 4d momentum $p^2$ is very small compared to the relevant scales ($R^{-1}$, $\Lambda$), one gets the following solution

\begin{equation}
\begin{array}{lcc}
\phi (r) \ = \ A \cdot \displaystyle{J_0(\frac{\mu}{\sqrt{\pi}}\ \frac{r}{\epsilon})} \quad \quad \quad \quad \quad \quad \quad \quad \quad \quad {\rm for} \quad r \leq \epsilon \ , \\
\phi(r) \ = \  B \displaystyle{\left[ J_0(|p|r)-\frac{J_0(|p|R)}{N_0(|p|R)}\cdot N_0(|p|r)\right]}  \quad  {\rm for} \quad r \geq \epsilon \ ,
\end{array}
\label{gensol}
\end{equation}
where $J_0$ and $N_0$ are Bessel functions of the first and second kind. The Dirichlet condition was imposed for the outside solution, and we assumed that $p^2$ could be negative.

Matching conditions for both the solution and its derivative $d \phi / d r$ at $r=\epsilon$ lead to the four-dimensional mass squared
\begin{equation}
 p^2 \equiv m_{(4)}^2 \ = \ \frac{8 \pi}{R^2} \ \frac{\mu_c^2 - \mu^2}{\mu_c^4} \ = \ \frac{8 \pi}{R^2} \ \frac{1}{\mu^2 (R^{-1})} \quad .
 \label{m4}
\end{equation}
 
This result can be understood qualitatively by noticing that the Dirichlet boundary condition
forces all modes to acquire a mass, in the spirit of the
Scherk--Schwarz mechanism \cite{ss}. Then from a naive 4d viewpoint,
the mass of a would-be zero mode has both the contribution due to
the boundary condition, and an additional contribution, of opposite
sign, coming from the localised tachyonic term, 
\begin{equation}
m^{2}_{(4),{\rm naive}} \ = \ \frac{z_0^2}{R^2} \ - \ \frac{\mu^2}{\pi R^2} \quad ,
\label{m1} 
\end{equation} 
where $z_0$ is the first zero of the Bessel function
$J_0$ and the factor $1 / (\pi R^2)$ in the second term comes from
the KK expansion of the zero mode. Equation (\ref{m1}) indeed
predicts phase transition, but for $\mu_c^2 = (\pi z_0^2)  $. The
correct value, eq.~(\ref{muc}), however, contains an inverse
logarithmic factor coming from the running and the correct
expression for the mass is actually slightly more involved than the
simple guess (\ref{m1}).\\

From equation (\ref{m4}), we conclude that exactly at the critical coupling, there indeed exists a massless mode. For $\mu^2$ slightly above $\mu_c^2$, the mode becomes tachyonic. In this region, as in the 4d Standard Model electroweak symmetry breaking, the $\lambda \phi^4$ term has to be taken into account, and the field develops expectation value. The classical six-dimensional solution in this case is
\begin{equation}
\phi_c^2 \ = \ \frac{\mu^2-\mu_c^2}{\lambda} \quad .
\end{equation}
The effective Higgs vev is therefore $\sigma^2=\int d^2 y \phi_c^2$.
Developping the field $\phi (r,p) = \phi_c(r) + \xi(r,p)$, one finds that the fluctuation's mass is
\begin{equation}
m_{\xi}^2 \ = \ \frac{16 \pi}{R^2}\  \frac{\mu^2 - \mu_c^2}{\mu_c^4} \ = \ 2 \ |m_{\rm tachyon}^2|\quad .
\end{equation}
This expression is valid for small $(\mu^2 - \mu_c^2)$.

\section{Low energy effective theory}

In this section, we compare the results of section \ref{6d} to the 4d effective theory. Specifically, we want to show that the 6d phase transition properties remain valid at low energy.

The interesting field configuration away from the brane is
\begin{equation}
    \phi(x^\mu, r) \ = \ \sigma (x^\mu) \cdot \zeta (r) \ ,
\label{eff1}
\end{equation}
where $\zeta(r)$ is the general outside solution (\ref{gensol}). It is normalised to unity $\int \zeta^2 d^2y = 1$, and therefore $\sigma$ has a canonical kinetic term. Resolving the brane is equivalent to defining $\zeta (0) = \zeta (\epsilon)$.

The effective 4d potential for $\sigma(x^{\mu})$ is obtained by integrating the potential (\ref{1}) and the transverse kinetic term over the extra dimensions. One finds
\[
  V_{eff} (\sigma) \ = \ \frac{m_{(4)}^2}{2} \sigma^2
+ \frac{\lambda_{(4)}}{4} \sigma^4 \ ,
\]
where
\[
\lambda_{(4)} = \frac{64 \pi^2}{\mu_c^8} \ \frac{\lambda}{R^4}.
\]

The value of $m_{(4)}^2$ is precisely the same as in (\ref{m4}),
which establishes the correspondence between the 6d and the 4d approaches
in the unbroken phase. For $m_{(4)}^2 <0$, the 4d expressions for
the vev and the Higgs mass are $\sigma^2  =  - m_{(4)}^2 / \lambda_{(4)}$ and $m_\xi^2 =  - 2 m_{(4)}^2$.
These coincide with the results of section \ref{6d}, so that the correspondence exists in the broken phase as well.

\section{Conclusions}

The toy model we presented in this small report possess rich physics: running couplings, phase transitions, spontaneous symmetry breaking and infrared strong dynamics which all occur at the level
of classical field theory. Interestingly enough, very light modes and standard four-dimensional physics are generated
out of a higher dimensional theory with large scales only. 
This mechanism could be of relevance for the problem of electroweak symmetry breaking and mass generation.
Of course, an important step to make before addressing this phenomenological issue is the
inclusion of other fields like chiral 4d fermions and gravitational interaction.
In these proceedings, we focused on the main basic features of the models under consideration. More detailed information can be found in Ref.~\cite{dpr}.

\section*{Acknowledgments}
I would like to express my gratitude to organising committee for very warm hospitality. This work was supported in part by the RTN grants MRTN-CT-2004-503369.


\section*{References}
\begin{thebibliography}{99}

\bibitem{gw}
W.D. Goldberger and M.B. Wise, \Journal{\PRD}{65}{025011}{2002} [arXiv:hep-th/0104170].
 
\bibitem{dgv}
E. Dudas, C. Grojean and S. K. Vempati, ``Classical running of neutrino masses from six dimensions'' [arXiv:hep-ph/0511001].

\bibitem{ss}
J. Scherk and J. H. Schwarz, \Journal{\NPB}{153}{61}{1979};\\
E. Cremmer, J. Scherk and J. H. Schwarz, \Journal{\PLB}{84}{83}{1979}

\bibitem{qm} 
R. J. Henderson and S. G. Rajeev, \textit{J.\ Math.\ Phys.} {\bf 39}, 749 (1998) [arXiv:hep-th/9710061 and arXiv:hep=th/9609109];\\
S. N. Solodukhin, \Journal{\NPB}{541}{461}{1999} [arXiv:hep-th/9801054].

\bibitem{dpr}
E. Dudas, C. Papineau, V. A. Rubakov, \textit{JHEP} {\bf 0603}, 085 (2006) [arXiv:hep-th/0512276]

\end{thebibliography}
\end{document}